\documentclass{appolb}

\usepackage{epsfig}
\usepackage{graphicx}

\newcommand{\Fi}[1]   {Fig.~\ref{#1}}

\newcommand{\agev}    {\mbox{$A$~GeV}}               

\newcommand{\rb}[1]   {\mbox{\textrm{\small #1}}}
\newcommand{\rbt}[1]  {\mbox{\textrm{\tiny #1}}}
\newcommand{\sqrts}   {\ensuremath{\sqrt{s_{_{\rbt{NN}}}}}}

\newcommand{\xim}     {\ensuremath{\Xi^{-}}}

\newcommand{\kmin}    {\ensuremath{\textrm{K}^-}}
\newcommand{\kplus}   {\ensuremath{\textrm{K}^+}}

\newcommand{\ommin}   {\ensuremath{\Omega^-}}
\newcommand{\pt}      {\ensuremath{p_{\rbt{t}}}}

\newcommand{\ptavg}   {\ensuremath{\langle p_{\rbt{t}} \rangle}}

\newcommand{\dnchdeta}{\ensuremath{\textrm{d}N_{\rbt{ch}}/\textrm{d}\eta}}

\newcommand{\nwound}  {\ensuremath{\langle N_{\rb{w}} \rangle}}

\newcommand{\tk}      {\ensuremath{T_{\rbt{kin}}}}
\newcommand{\tc}      {\ensuremath{T_{\rbt{C}}}}
\newcommand{\tch}     {\ensuremath{T_{\rbt{ch}}}}
\newcommand{\mub}     {\ensuremath{\mu_{\rbt{B}}}}

\newcommand{\btavg}   {\ensuremath{\langle \beta \rangle}}

\newcommand{\tauf}    {\ensuremath{\tau_{\rbt{f}}}}

\newcommand{\kstar}   {\ensuremath{\textrm{K}^{*}\textrm{(892)}}}
\newcommand{\kstarbar}{\ensuremath{\bar{\textrm{K}}^{*}\textrm{(892)}}}
\begin{document}

%
\title{Is there Life after Hadronization? \\
An Experimental Overview.%
\thanks{Presented at the conference ``Strangeness in Quark Matter
  2011'' in Cracow.}%
}

%
\author{Christoph Blume
\address{Institut f\"ur Kernphysik der J.W.~Goethe Universit\"at, \\
60438 Frankfurt am Main, Germany}
}
\maketitle

%
\begin{abstract}
Recents experimental findings on the properties of the chemical and
kinetic freeze-out are reviewed, including data from low energies
(SPS) over RHIC, up to recent results from the LHC.  We discuss
whether chemical freeze-out coincides with hadronization or if there
is evidence for a "life after hadronization" which might significantly
change particle abundances.
\end{abstract}
  
%
\section{Introduction}

The conventional and simplified scenario of the evolution of a high
energy heavy ion reaction is the following: Starting from a
Quark-Gluon Plasma (QGP) phase particles hadronize at the critical
temperature \tc\ and the system then turns into a hadron gas which
further evolves as it expands and cools down.  It will reach the
chemical freeze-out temperature \tch\ at which all inelastic reactions
cease and the relative particle abundances are frozen out.  Afterwards
elastic scattering processes may still continue and possibly modify
the momentum distributions of the hadrons until the kinetic freeze-out
temperature \tk\ is reached.  After this final freeze-out only free
streaming particles propagate towards the detector.  The question to
be addressed in the following is whether there is any experimental
indication for these different phases after hadronization and, if yes,
if there is any way of determining the lifetimes of the different
phases by measurements.

It is by no means clear that the above described scenario is realized
in nature.  In fact, there are several other concepts that do not
require any extended period between hadronization and chemical or
kinetic freeze-out.  Single freeze-out models, for instance, where
chemical and kinetic freeze-out coincide, have been quite successful
in describing particle spectra and yields as measured at RHIC
\cite{BRONIOWSKI01}.  The suggestion that chemical equilibration is
generally driven by the vicinity of a phase boundary implies that
there is not need for an extended hadron gas phase between
hadronization and chemical freeze-out \cite{PBM11}.  The same applies
to sudden freeze-out models as discussed in \cite{RAFELSKI02}.

One way to model an extended hadronic phase after hadronization in
theory is to couple a hadronic transport model as an afterburner to a
hydrodynamic calculation \cite{BASS00}.  Alternatively, a
hydrodynamical evolution has been combined with an extended freeze-out
description \cite{KNOLL09} to investigate the effect of an extended
chemical freeze-out.  These approaches allows to study whether
particle abundances and spectra are modified by subsequent hadronic
interactions as, e.g., the absorption of anti-baryons.  As a
consequence one might observe deviations from the chemical equilibrium
value for certain particle species.    It has also been pointed out
quite early that particles with low hadronic cross sections
(e.g. \ommin, $\phi$) should freeze-out earlier than other hadrons if
there was a long lived hadronic phase \cite{HECKE98}.

%
\section{Kinetic Freeze-Out}

%
\begin{figure}[t]
\begin{center}
\begin{minipage}[b]{0.49\linewidth}
\begin{center}
\includegraphics[width=\linewidth]{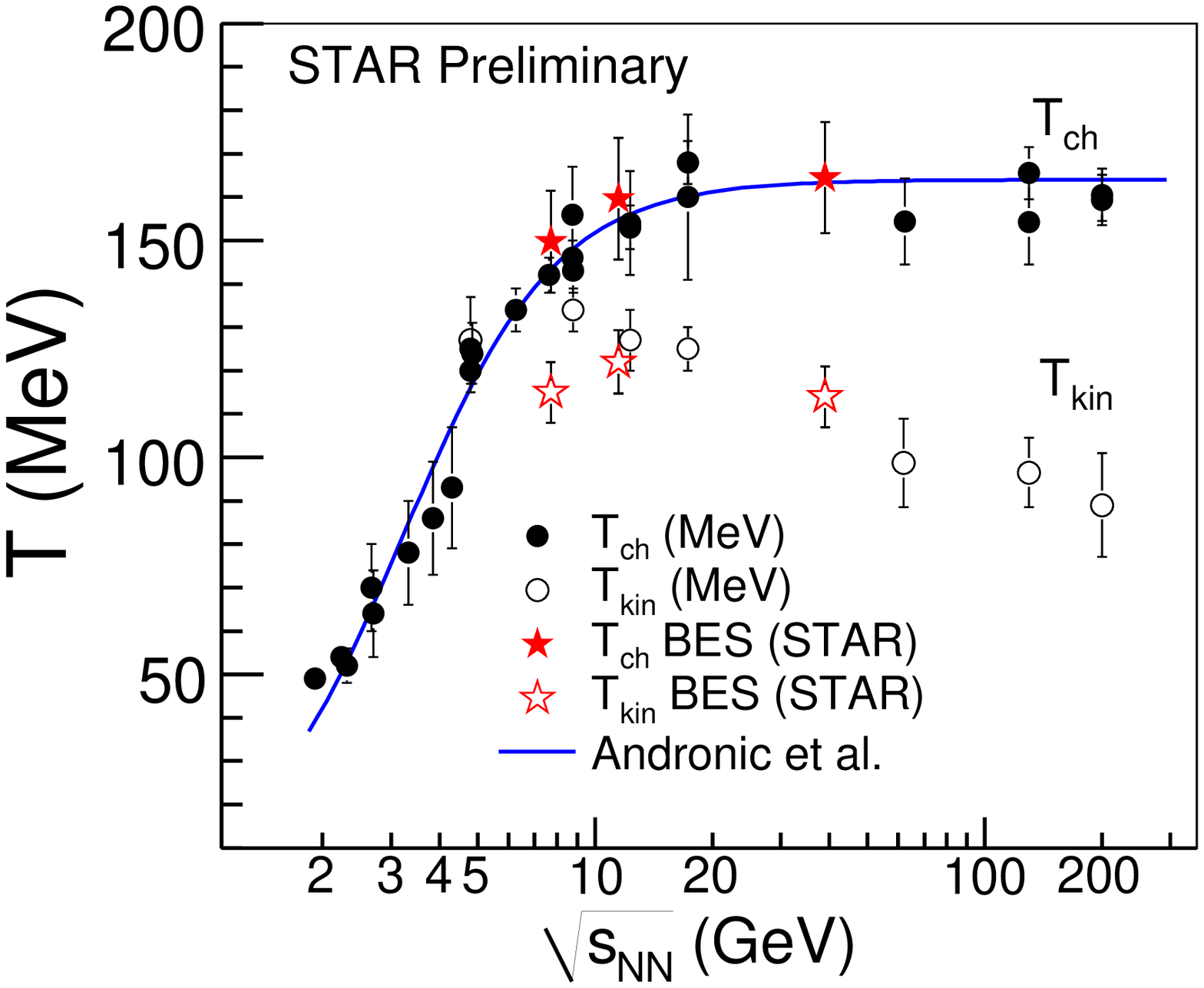}
\end{center}
\end{minipage}
\begin{minipage}[b]{0.49\linewidth}
\begin{center}
\includegraphics[width=0.94\linewidth]{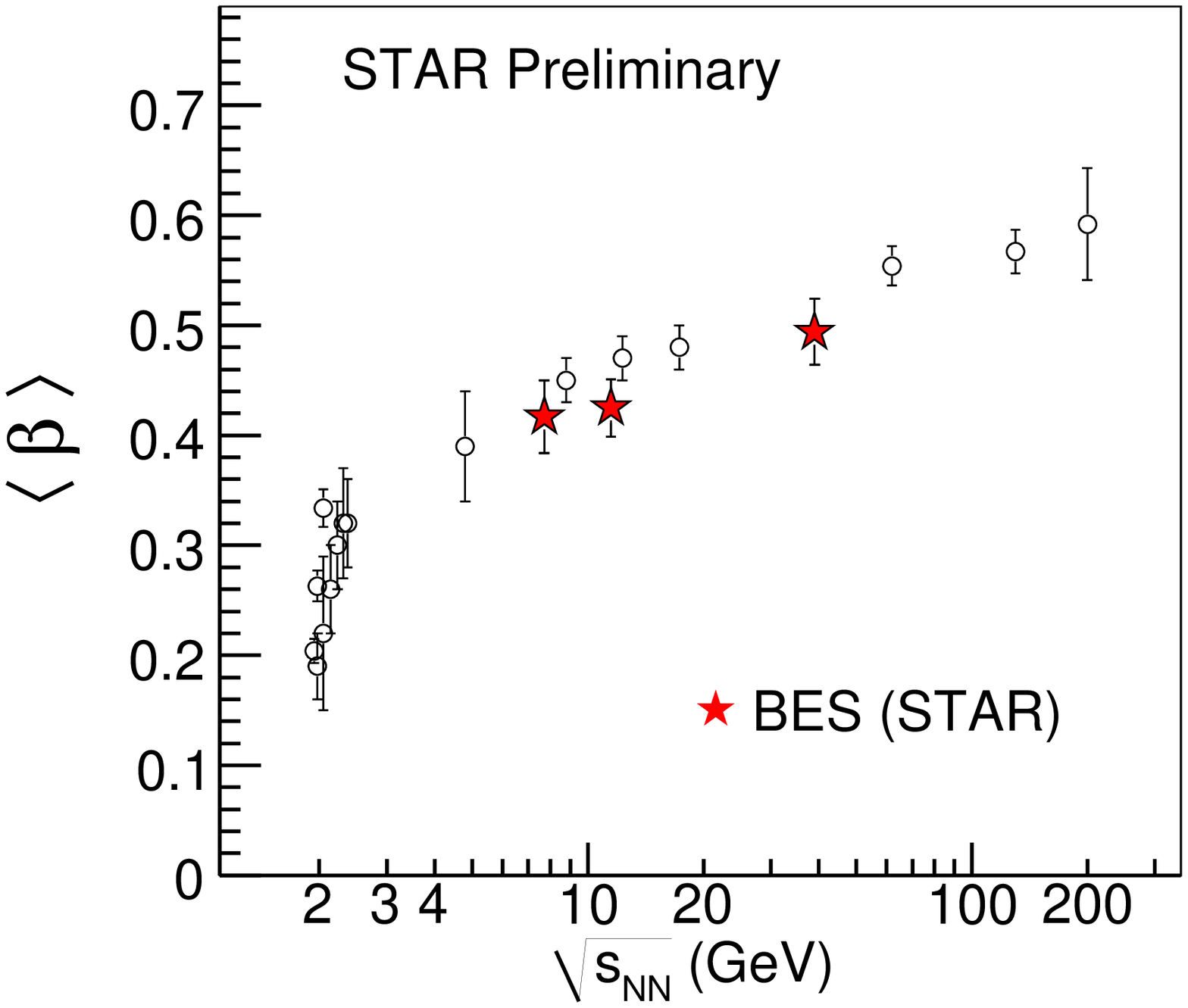}
\end{center}
\end{minipage}
\end{center}
\caption{Left: The chemical \tch\ (filled symbols) and kinetic \tk\
  (open symbols) freeze-out temperatures as a function of \sqrts\
  \cite{KUMAR11}.  The solid line shows a parametrization of \tch\
  \cite{ANDRONIC06}.
  Right: The mean transverse expansion velocity \btavg\ as a function
  of \sqrts\ \cite{KUMAR11}.}
\label{fig:bw_star}
\end{figure}
%

The left panel of \Fi{fig:bw_star} shows a recent compilation of
chemical \tch\ and kinetic \tk\ freeze-out temperatures as a function
of \sqrts\ \cite{KUMAR11}.  \tch\ has been extracted from fits with
statistical models (e.g. \cite{ANDRONIC06,BECATTINI01,WHEATON09}) to
particle yields measured in central nucleus-nucleus collisions at
different energies, while \tk\ was determined by fits to transverse
momentum spectra of pions, kaons, and protons with a blast wave model
\cite{SCHNEDERMANN94}.  The latter model provides a simple
parametrization of \pt~spectra which includes the effect of transverse
expansion, thus allowing a simultaneous fit to the spectra of hadrons
with different mass with usually reasonable results.  The right panel
of \Fi{fig:bw_star} shows the resulting average transverse expansion
velocity \btavg.  There is a clear difference between \tch\ and \tk\
seen in these fit results with $\tk < \tch$.  While \tch\ reaches a
plateau value from about \sqrts~= 10~GeV onwards, \tk\ rather
decreases with increasing \sqrts.  This finding is corroborated by
resent results from ALICE which indicate an even lower \tk\ than
observed at top RHIC energies \cite{FLORIS11}.  The natural
explanation is an extended hadronic phase between chemical and
kinetic freeze-out whose lifetime is increasing as the center-of-mass
energy and thus the initial energy density of the system is
increasing.  This finding agrees with the observation that the
freeze-out volume extracted from HBT radii is larger by a factor of
two at LHC compared to RHIC \cite{ALICEHBT}.  From the LHC data
freeze-out times of \tauf~= 10~-~11~fm/$c$ have been extracted, which
are 40~\% larger than what is seen at RHIC.

%
\begin{figure}[t]
\begin{center}
\includegraphics[width=\linewidth]{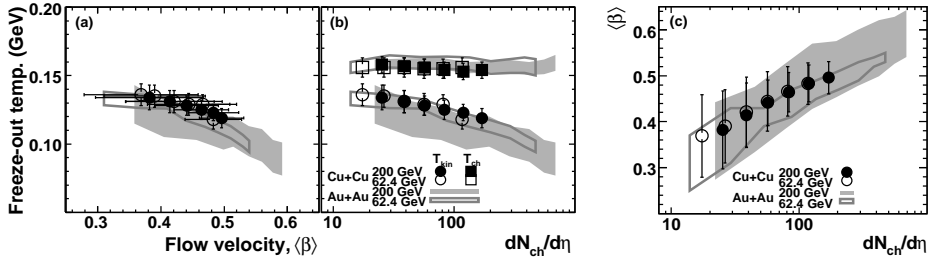}
\end{center}
\caption{The centrality dependence of chemical and kinetic freeze-out
  parameters for Cu+Cu and Au+Au collisions at \sqrts~= 62.4 and
  200~GeV \cite{STARSIZE}.  Shown is \tk\ versus \btavg\ (a), as well
  as \tk\ and \tch\ (b), and \btavg\ (c) all three as a function of
  \dnchdeta.}
\label{fig:tkin_star}
\end{figure}
%

To address the question whether freeze-out happens due to a dynamic
decoupling from interactions in a hadronic stage, the study of the
system size dependence of freeze-out parameters plays a decisive role.
Due to the interplay between the mean free path of the hadronic
scatterings and the changing size of the system, one expects a clear
change of freeze-out temperatures if the system size varies
\cite{HEINZ06}.  In fact, this is being observed for the kinetic
freeze-out temperature (see \Fi{fig:tkin_star},b).

%
\begin{figure}[ht]
\begin{center}
\includegraphics[width=0.8\linewidth]{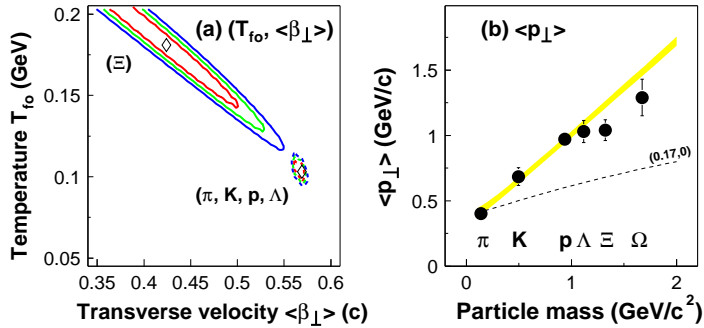}
\end{center}
\caption{Left: The kinetic freeze-out temperature \tk\ and average
  transverse expansion velocity \btavg\ extracted from blast wave fits
  to light hadrons ($\pi$, K, p, $\Lambda$) and the $\xim$ for Au+Au
  collisions at \sqrts~= 200~GeV \cite{STARBW}.
  Right: The average transverse momentum \ptavg\ for different hadrons
  measured in Au+Au collisions at \sqrts~= 200~GeV \cite{STARBW}.}
\label{fig:bw_hyperon}
\end{figure}
%

Figure~\ref{fig:bw_hyperon} shows a comparison of the kinetic
freeze-out conditions of light hadrons ($\pi$, K, p, $\Lambda$) and
the rare \xim\ and \ommin\ as measured at RHIC.  As the left panel
illustrates, there are clear indications from the blast wave fits that
\xim\ freeze-out at a higher temperature \tk.  Similarly, the average
transverse momenta \ptavg\ of \xim\ and \ommin\ do not follow the
roughly linear increase with particle mass as observed for lighter
hadrons.  The same has been concluded from data taken at the SPS (for
a recent review see \cite{BLUME11}).  This would support the picture
that rare particles with low hadronic cross section do decouple
earlier from an interacting hadronic phase and are thus less affected
by its transverse expansion \cite{XU02}.

On the other side, single freeze-out models ($\tch = \tk$) provide a
very good description of not only the light hadron \pt~spectra
measured at RHIC \cite{FLORKOWSKI05}, but also of the ones of
multi-strange hyperon (\xim\ and \ommin) \cite{BARAN04}.  In contrast
to the previous conclusions this would indicate that there is no need
for an extended hadronic phase between chemical and kinetic freeze-out
to explain the differences between the different hadron species.  In
these models the origin of the difference is rather that the spectra
of light particles are strongly affected by resonance decays, which
is not the case for particles as the \ommin\ or $\phi$.

%
\begin{figure}[ht]
\begin{center}
\begin{minipage}[b]{0.49\linewidth}
\begin{center}
\includegraphics[width=0.94\linewidth]{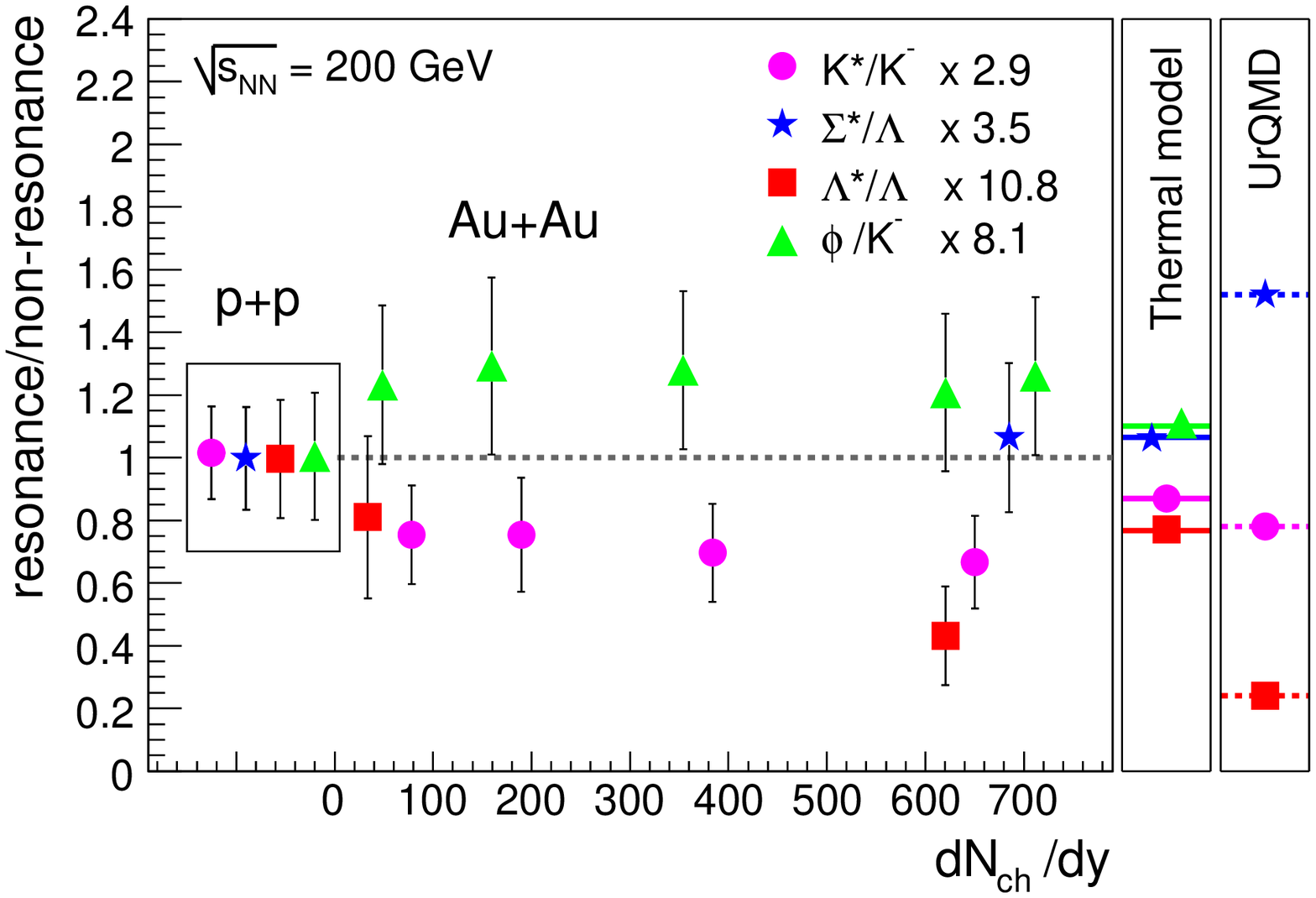}
\end{center}
\vspace{1mm}
\end{minipage}
\begin{minipage}[b]{0.49\linewidth}
\begin{center}
\includegraphics[width=\linewidth]{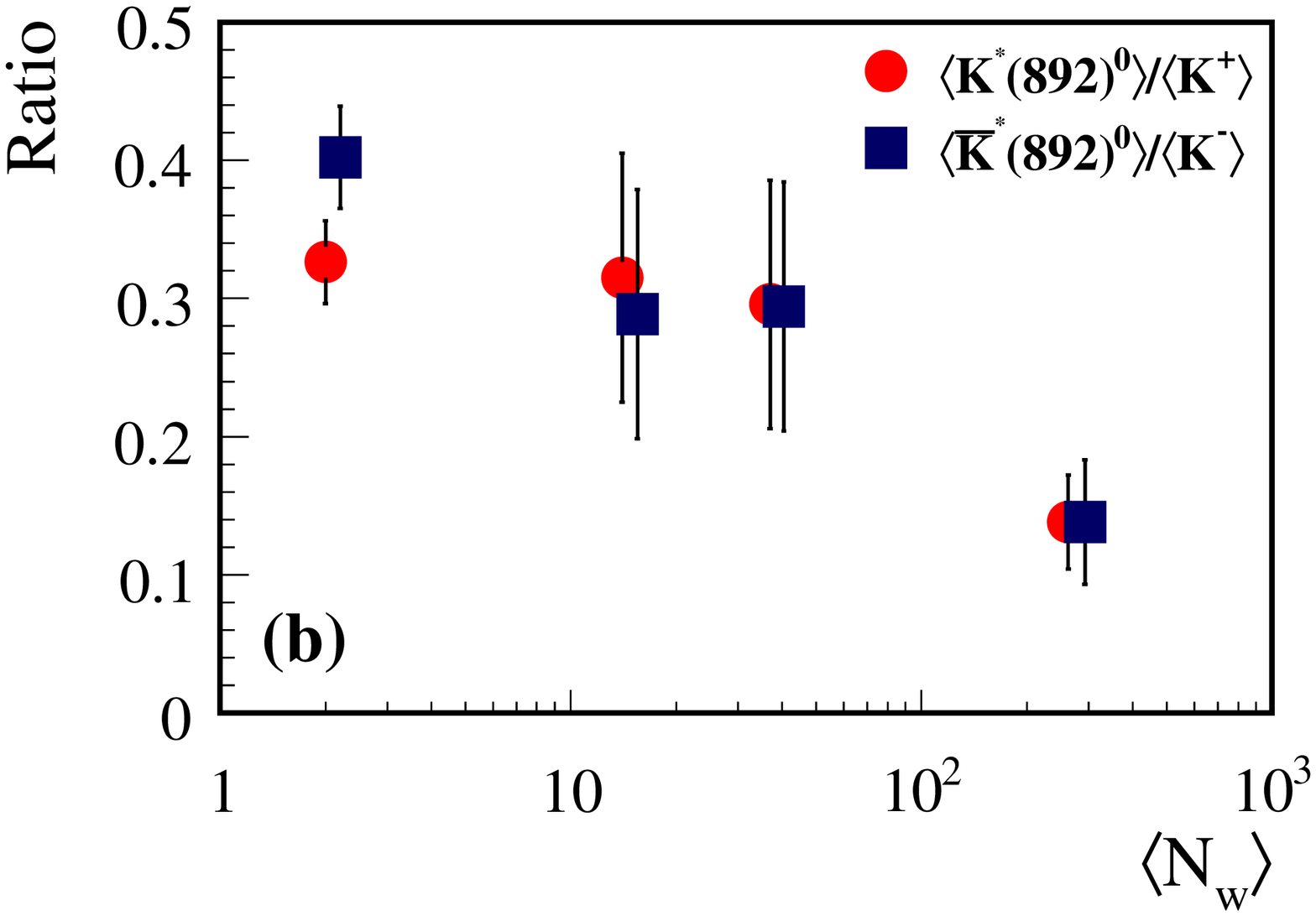}
\end{center}
\end{minipage}
\end{center}
\caption{Left: Resonance to stable particle ratios for p+p and
Au+Au collisions at \sqrts~= 200~GeV \cite{STARRES}. The ratios are
normalized to unity in p+p.
Right: The total yield of \kstar\ (\kstarbar) divided by the total
yields of \kplus\ (\kmin) in p+p and nucleus-nucleus collisions at
158\agev\ as a function of the average number of wounded nucleon
\nwound \cite{NA49RES}.}
\label{fig:resonances}
\end{figure}
%

More informations on the lifetime of a phase between chemical and
kinetic freeze-out can possibly be derived from the measurement of
resonance yields.  Since strange resonances such as the \kstar\ and
$\Lambda$(1520) have a lifetime that is comparable to the lifetime of
the fireball, they will decay to a certain fraction inside of it.  As
a consequence the momenta of their decay products could be modified by
elastic scattering processes and thus make the experimental
reconstruction of the resonance itself via an invariant mass analysis
impossible.  One would therefore measure less resonances than expected
from statistical model predictions.  In fact, experimental
observations are in accordance with this scenario.  The ratio of
resonance to non-resonance particles (e.g. \kstar/\kmin) shows a
decrease with increasing system size (see \Fi{fig:resonances}), as
expected due to the increasing rescattering time before kinetic
freeze-out.  However, a quantitative analysis will require also more
accurate data.  Right now it seems that the effect is even stronger at
SPS than at RHIC which would be in contradiction to the larger
freeze-out times extracted at higher energies.  Upcoming data from LHC
will shed more light on this issue.

%
\section{Chemical Freeze-Out}

%
\begin{figure}[ht]
\begin{center}
\begin{minipage}[b]{0.49\linewidth}
\begin{center}
\includegraphics[width=0.75\linewidth]{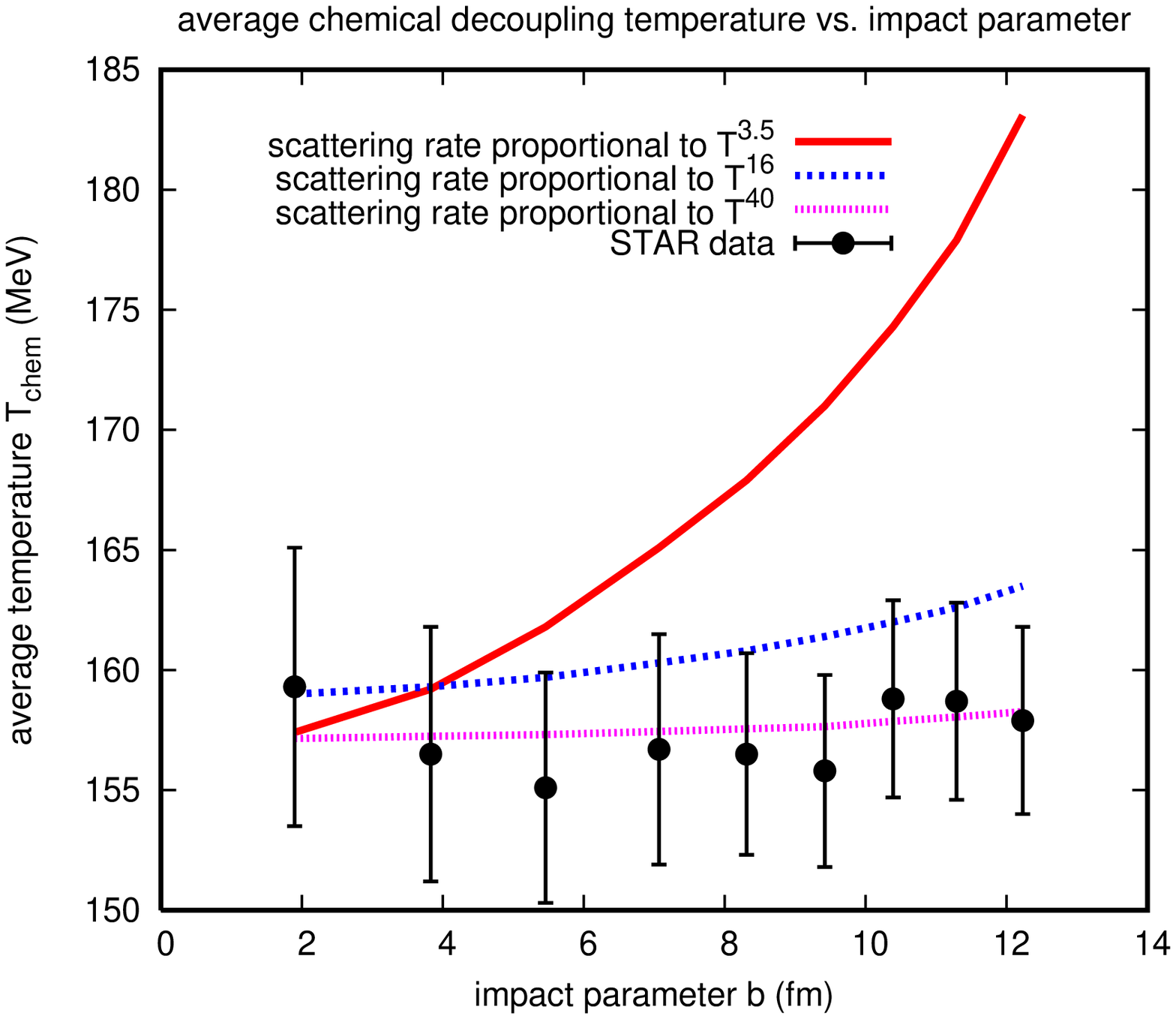}
\end{center}
\vspace{1mm}
\end{minipage}
\begin{minipage}[b]{0.49\linewidth}
\begin{center}
\includegraphics[width=\linewidth]{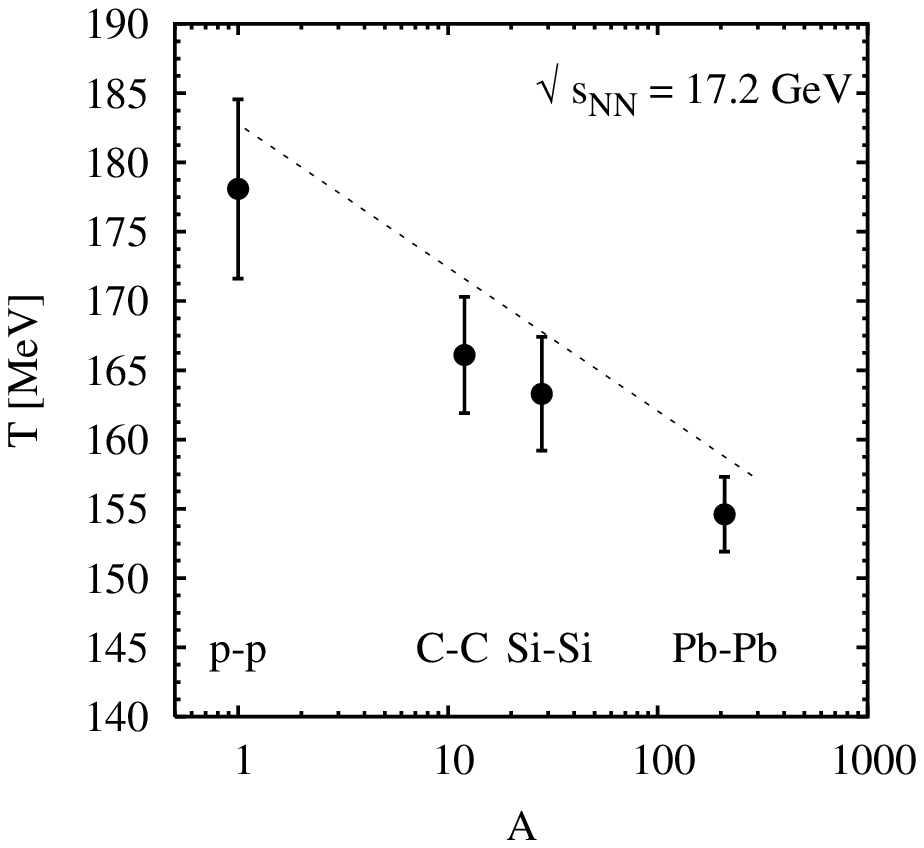}
\end{center}
\end{minipage}
\end{center}
\caption{Left: The impact parameter dependence of the average chemical
  decoupling temperature \tch\ computed from hydrodynamics with
  different temperature dependences of the scattering rates
  \cite{HEINZ06}, compared with STAR data for Au+Au collisions at
  \sqrts~= 200~GeV \cite{STARSIZE2}.
  Right: The fitted temperature at chemical freeze-out \tch\ as a
  function of the mass number $A$ in central heavy ion collisions at
  \sqrts~= 17.3~GeV (from left to right: p+p, C+C, Si+Si, and Pb+Pb).
  The dashed line shows a parametrization \cite{BECATTINI06}.}
\label{fig:syschem}
\end{figure}
%

The question whether there is any evidence for an extended hadronic
phase after hadronization and before chemical freeze-out is much more
difficult to answer.  One indicator might again be the production of
particles with low hadronic cross sections like multi-strange baryons.
However, there is no observation yet that any of these particles shows
a significant deviation from its statistical equilibrium value as
expected in the case of an early hadronic decoupling \cite{BLUME11}.
However, the error bars on these data, especially at lower energies,
are still relatively large so that for a more definite statement also
improved data sets would be needed.  On the other side, it has also
been argued that multi-strange particles might not be too sensitive to
a hadronic phase, if they already enter this phase with abundances
close to equilibrium \cite{BASS00}.  Antibaryons are expected to be
much stronger affected due to absorption effects which would cause a
reduction of the measured yields relative to the chemical equilibrium
value. A caveat here is that antibaryon production can be enhanced in
a hadron gas by multi-meson fusion processes \cite{RAPP01}, which
usually are not included in hadronic transport models.  A recent study
of statistical model fits to data sets including and excluding
antibaryons nevertheless indicates that there might be a significant
effect \cite{STOCK11}.  Also, antiprotons exhibit a decrease of their
measured yields with increasing system size \cite{STARSIZE,NA49PBAR}
as expected if they were subject to increasing absorption in hadronic
matter.  However, this needs to be disentangled from the system size
evolution of the effects of baryon number transfer to mid-rapidity
that could result in a similar observation.  A striking observation
that was made recently when results on particle yields for Pb+Pb
collisions at \sqrts~= 2.76~TeV were compared to statistical model
predictions was that the measured proton and antiproton yields are
significantly lower than the expectations \cite{FLORIS11}.  However,
whether this might be an indication for hadronic rescattering in long
lived hadron gas phase will still need further investigations.

One remarkable observation made at RHIC is that the system size
dependence of \tch\ and \tk\ are clearly different (see
\Fi{fig:tkin_star},b).  While \tk\ shows a significant dependence on
system size, \tch\ does not.  As pointed out in \cite{HEINZ06}, this
is very difficult to be reconciled with a freeze-out from a normal
hadron gas.  To describe this behavior scattering rates that depend
on temperature as $T^{n}$ with $n > 20$ are required (see left panel
of \Fi{fig:syschem}).  It has been pointed out in \cite{PBM04} that
exponents of the order of 60 can be realized if chemical freeze-out
happens directly at the phase boundary.  At high energies and small
\mub\ the vicinity deconfinement phase transition would thus provide
an explanation for the observed feature.  Since at lower energies and
consequently higher \mub\ the system freeze-out far away from this
phase boundary, it is important to do a systematic investigation of
the system size dependence of chemical freeze-out parameters also
here.  One example is shown in the right panel of \Fi{fig:syschem}.
Here a clear $A$~dependence of \tch\ is found \cite{BECATTINI06},
which is in stark contrast to the observations at RHIC and could be
interpreted as an effect of hadronic re-interaction.  However, one has
to take into account that simple geometric effects, as expected in the
core-corona scenario \cite{BOZEK05,BECATTINI08,AICHELIN09}, can play a
dominant role \cite{BLUME10}.  At SPS energies \tch\ is found to be
quite different between p+p and A+A, which is not the case at top RHIC
energy \cite{BECATTINI06}.  Therefore a core-corona superposition will
lead in the first case to a system size dependence of \tch, in the
second case not.  Recently, new data from the beam energy scan program
of STAR were shown \cite{KUMAR11-2}.  Also here a system size
dependence of \tch\ was reported for heavy ion collisions at lower
energies (\sqrts~= 7.7, 11.5, and 39.0~GeV).  However, in contrast to
what is shown in \Fi{fig:syschem} and what one would expect if smaller
systems freeze-out earlier and thus at higher temperatures, a decrease
of \tch\ towards small systems is observed.

%
\section{Conclusions}

The answer to the initial question ``Is there life after
hadronization?'' turns out to be not straight forward.  Seen from a
phenomenological standpoint all observations (\tk\ from blast wave
fits, \tk\ depends in system size, low cross section particles seem
to freeze-out earlier, resonances get suppressed) are in accordance
with the existence of a long lived hadronic phase between chemical and
kinetic freeze-out.  However, there is also the remarkable success of
single freeze-out models which invalides many of the usual arguments.
For the existence of a hadronic phase before chemical freeze-out no
clear evidence can be derived from the current experimental data.
However, there are several not fully understood features (antiprotons
at LHC, system size dependence of \tch\ at low energies) which might
provide some clue for future investigations.

%
\section*{Acknowledgements}

The author would like to thank the organizers for the kind invitation
to this interesting and stimulating conference and acknowledges
helpful discussions with F.~Becattini, M.~Bleicher, M.~Ga\'{z}dzicki,
C.~Markert, G.~Torrieri, T.~Schuster, J.~Stachel, and R.~Stock.

%

\end{document}